\documentclass[preprint,showpacs,preprintnumbers,amsmath,amssymb,apsrev,aps,prb]{revtex4}
\usepackage{graphicx}% Include figure file
\usepackage{dcolumn}% Align table columns on decimal point
\usepackage{pstricks}
\usepackage{epsf}
\usepackage{bm}% bold math
\usepackage[latin1]{inputenc}
\newcommand{\comment}[1]{}
\newcommand\etal{\mbox{\textit{et al.~}}}

\begin{document}
\setlength{\unitlength}{0.7\textwidth}
%\preprint{}

\title{The role of coherent vorticity in turbulent transport in resistive drift-wave turbulence}

\author{W.J.T. Bos$^{1,2}$, S. Futatani$^{3,4}$, S. Benkadda$^3$, M. Farge$^5$ and   K. Schneider$^1$ 
}
\affiliation{ 
$^1$ M2P2 -- CNRS \& CMI, Universit\'e de Provence, 39, rue Joliot-Curie
13453 Marseille Cedex 13, France\\
$^2$ LMFA -  UMR 5509, - CNRS - Ecole Centrale de Lyon - Universit\'e Claude
Bernard Lyon 1 - INSA de Lyon, 69134, Ecully Cedex, France\\
$^3$ France-Japan Magnetic Fusion Laboratory LIA 336 CNRS / UMR 6633, CNRS-Universit\'{e} de Provence. Case 321, 13397 Marseille Cedex 20, France\\ 
$^4$ Graduate School of Energy Science, Kyoto University, Japan\\
$^5$ LMD--CNRS, Ecole Normale Sup\'erieure Paris, 24 rue Lhomond, 75231 Paris cedex 05, France
}

\date{\today}% It is always \today, today,
%\date{June 6, 2008}

\begin{abstract} 
The coherent vortex extraction method, a wavelet technique for extracting coherent vortices out of turbulent flows, is applied to simulations of resistive drift-wave turbulence in magnetized plasma (Hasegawa-Wakatani system). The aim is to retain only the essential degrees  of freedom, responsible for the transport. It is shown that the radial density flux is carried by these coherent modes. In the quasi-hydrodynamic regime, coherent vortices exhibit depletion of the polarization-drift nonlinearity and vorticity strongly dominates strain, in contrast to the quasi-adiabatic regime.
\end{abstract}

\maketitle

\section{Introduction and governing equations}

One important issue in fusion research is the understanding and
 control of turbulent radial flux of particles and heat in magnetized
 plasmas, in order to improve the confinement properties of fusion
 devices\cite{Garbet2006}.
%{\bf
%Add reference 
%}
 Indeed turbulence enhances the radial diffusion dramatically compared
 to neo-classical estimations. A long standing question has been
 \cite{Zweben1985,Koniges1992,Dudok1995,Horton1996,Hu1997}: what is
 the role of coherent structures in this radial transport? The answer
 to this question 
%cannot be given without 
requires
extracting and characterizing coherent structures. 
A particularly appropriate framework to identify coherent structures
 is the wavelet representation, 
%a representation which is well
where wavelets are basis functions well
localized in both physical and Fourier space \cite{Farge92}. 
It has already been used 
%The application of the wavelet decomposition has revealed its capacity 
to identify coherent structures in fluid turbulence and to
 distinguish them from background incoherent noise
 \cite{Pellegrino2001}. These methods have recently been applied to
 experimental signals of ion density in the tokamak scrape-off layer
 \cite{Farge2006}, separating coherent bursts from incoherent
 noise. In the present work these methods are applied  to assess the
 role of coherent vorticity structures in anomalous radial transport
 in two-dimensional numerical simulations of drift-wave turbulence. 
%{
Drift waves are now generally considered to play a key role in
  the dynamics and transport properties of tokamak edge turbulence
  (e.g. [\onlinecite{Scott2002}] and references therein). At the edge,
  the plasma temperature is low and the collision rate relatively
  large, therefore the resistivity is potentially important. The
  Hasegawa-Wakatani model\cite{Hasegawa1983,Wakatani1984} is a
  two-field model which includes the main features of turbulent transport by resistive drift waves.
%}

%To study plasma-edge turbulence, Hasegawa and Wakatani
% proposed a two-field model which
%includes the main features of turbulent transport. 
In the present work
the two-dimensional slab geometry-version of this model is chosen as a
paradigm for drift-wave turbulence in the plasma-edge region.
In dimensionless form the Hasegawa-Wakatani model  reads \cite{Horton1990}
\begin{eqnarray}
\left(\frac{\partial}{\partial t}-D\nabla^2 \right)n+\kappa\frac{\partial \phi}{\partial y}+c(n-\phi)=\left[n,\phi\right],\label{hw1}\\
\left(\frac{\partial}{\partial t}-\nu\nabla^2 \right)\nabla^2 \phi+c(n-\phi)=\left[\nabla^2 \phi,\phi\right],\label{hw2}
\end{eqnarray}
%{\bf 
with $n$ the plasma density fluctuation and $\phi$ the electrostatic
potential fluctuation. $D$ and $\nu$ are the cross-field diffusion of plasma density fluctuations and kinematic viscosity, respectively. The
Poisson brackets are defined as 
\begin{equation}
[a,b]=\frac{\partial a}{\partial x}\frac{\partial b}{\partial y}-\frac{\partial
a}{\partial y}\frac{\partial b}{\partial x}.
 \end{equation}
We identify the $x$-coordinate with the radial direction and the
$y$-coordinate with the poloidal direction. The equilibrium density
$n_0$ is non-uniform, with a density gradient $dn_0/dx$ in the
negative $x$-direction, such that the equilibrium density scale
$L_n=n_0/(dn_0/dx)$ is constant and the value of $\kappa$ is one. The plasma density fluctuations $n$ are normalized  by $n_0$, therefore $n/n_0\rightarrow n$, the electrostatic potential is normalized as $e\phi/T_e\rightarrow\phi$, the space as $x/\rho_s\rightarrow x$ and the time as $\omega_{ci}t\rightarrow t$, where $e$ is the electron charge, $T_e$ the electron temperature,  $\omega_{ci}$ the ion cyclotron frequency and $\rho_s=(m_iT_e)^{1/2}/(eB)$ is the ion integral
lengthscale. $B$ is the strength of the equilibrium magnetic field in the $z$-direction and $m_i$ is the ion mass. The key parameter in this model is the adiabaticity $c$, which represents the strength of the parallel electron resistivity. It is defined as 
\begin{equation}
c=\frac{T_e k_{\parallel}^2}{e^2n_0\eta\omega_{ci}},
\end{equation}
with $k_{\parallel}$ the effective
parallel wavenumber and $\eta$ the electron
resistivity.
%}

The vorticity $\omega$ is related to the electrostatic potential
$\phi$ by 
\begin{equation}
\nabla^2 \phi=\omega.
\end{equation}  
Note that for $c=0$, equation (\ref{hw1}) corresponds to the advection-diffusion of a passive scalar in the presence of a (unity) mean scalar gradient in the
$x$-direction. Equation (\ref{hw2}) corresponds in this case to the
vorticity equation. For  $c\rightarrow \infty$ the Hasegawa-Mima \cite{Hasegawa1977} one field  approximation  is approached,\cite{Horton1999}
which ignores all resistive effects. For $c\rightarrow 0$ we recover
the hydrodynamic limit, which is less relevant to describe edge
fusion-plasma. Here two cases will be considered: a quasi-adiabatic
case with $c=0.7$, and a quasi-hydrodynamic case with $c=0.01$. 
%{\bf
The case $c=0.7$ is generally considered to be the most relevant
 for tokamak-research and has been investigated in several other works (e.g. [\onlinecite{Horton1999,Dudok1995}]).
%}
Both cases differ from the fluid-dynamical case in that the velocity field is forced through the interaction term $c(n-\phi)$. The influence of this term on the density field can however be considered to be negligible in the quasi-hydrodynamic case \cite{Hu1997}.

\begin{figure}
\setlength{\unitlength}{1.\textwidth}
\includegraphics[width=0.5\unitlength]{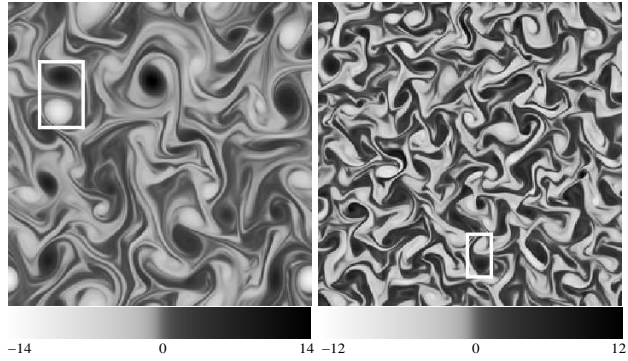}
\caption{One realization of the vorticity field for the
  quasi-hydrodynamic case (left) and for the quasi-adiabatic case
  (right). 
%{\bf 
The abscissa corresponds to the radial position. The
  ordinate indicates the poloidal position. Both range from 0 to 64 $\rho_s$. The white frames indicate the dipoles we have selected in both cases.
%}
\label{VortFields}}
\end{figure}

 The quantity of interest, the radial particle density flux, is the correlation between the radial velocity $u_r=-\partial \phi/\partial y$ and the particle density,
\begin{equation}
\Gamma_r=\left<n u_r\right>,
\end{equation}
where the brackets denote an average over both time and space. The question
 we address in this paper is how coherent structures contribute to
 this flux. To investigate this, 
%computations are performed on a
%periodic domain discretized with $N=512^2$ gridpoints.
%{\bf
direct numerical simulations of the Hasegawa-Wakatani system are performed on a
periodic domain discretized with $N=512^2$ gridpoints. The length of
the domain is 64 $\rho_s$. A finite difference method is used in which
the nonlinear terms are computed using a method developed by Arakawa
\cite{Arakawa1966}. The time stepping is performed using a
predictor-corrector scheme. 
%}
The plasma density diffusion $D$ and viscosity $\nu$ are set to $0.01$ in normalized units.  Computations are performed up to $t=612$. At $t\approx100$ the kinetic energy saturates and a statistically stationary state is reached, independent of the (random) initial conditions. Typical realizations of the vorticity field are shown in figure \ref{VortFields}, where one observes coherent structures for both cases. In each case we select a dipolar structure that we indicate by a white frame. The quasi-hydrodynamic case exhibits coherent vortices of very different sizes and intensities, in contrast to the quasi-adiabatic case where the coherent structures are more similar in size and intensity.

\section{Coherent Vortex Extraction (CVE)}

\subsection{Method}

%{\bf
Definitions and details on the orthogonal wavelet transform and its
extension to higher dimensions can be found, e.g., in
[\onlinecite{Farge92,Mallat98}].
In the following we fix the notation for the orthogonal
wavelet decomposition of a two--dimensional scalar valued field. The
wavelet transform unfolds the field into scales, positions and
directions using a set of dilated, translated and rotated functions, called wavelets. Each wavelet is
well-localized in space, 
%({\it {\it i.e.},} it exhibits a fast decay
%for $|\vec x|$ tending to infinity), 
oscillating ({\it {\it i.e.},} it has at least a vanishing mean, or better its first $m$ moments vanish), and smooth ({\it {\it i.e.},} its Fourier
transform exhibits fast decay for wavenumbers tending to infinity). 
%The mother wavelet then generates a
%family of wavelets $\psi_{\lambda} (\vec x)$, by dilatation, translation and rotation. 
We here apply the coherent vortex extraction (CVE) algorithm \cite{FSK99,Pellegrino2001} using  orthogonal wavelets. In dimension two, orthogonal wavelets span
three directions (horizontal, vertical and diagonal), due to the tensor product construction. To go from one scale to the next, wavelets are dilated by a factor two and the translation step doubles accordingly. Wavelet coefficients are thus represented on a dyadic grid\cite{Farge92}.

%The size of the wavelets is varied by dyadic dilatation such that the size of the wavelets of successive scales differs by a factor two. 

We apply the CVE algorithm to the vorticity fields $\omega$
of both the quasi-hydrodynamic and
the quasi-adiabatic regime.  The
extraction is performed from the vorticity since enstrophy is an
inviscid invariant in the hydrodynamic limit. Moreover, vorticity is
Galilean invariant in contrast to velocity and streamfunction. We
consider the quasi-stationary state of the simulations, i.e., when a
saturated regime is reached, and we decompose the vorticity field, given at resolution $N=2^{2J}$, into an orthogonal wavelet series
%}

%

%We apply the coherent vortex extraction (CVE) algorithm
%\cite{Pellegrino2001} to both cases, using the Coifman 30 orthogonal
%wavelet, which has 10 vanishing moments. The extraction is performed
%from the vorticity field since enstrophy is an inviscid invariant in
%the hydrodynamic limit. Moreover, vorticity is Galilean invariant in contrast to velocity and streamfunction. After reaching the saturated,
%quasi-stationary phase of the simulations, the vorticity field, 
%is projected onto a wavelet basis. 
%The strongest wavelet coefficients, whose modulus is above a
%threshold which depends on the resolution $N$ and enstrophy $Z$, are
%retained and considered to be part of the coherent vorticity field
%\cite{Pellegrino2001}. The incoherent vorticity is reconstructed from
%the remaining weak wavelet coefficients. 
%given at resolution $N=2^{2J}$, is projected onto a wavelet basis, 
%

\begin{equation}
\omega(x,y) = \sum_{\lambda \in \Lambda} \widetilde \omega_\lambda \psi_\lambda (x,y),
\end{equation}
where the multi--index $\lambda= (j,i_x,i_y,d)$ denotes the scale $j$ the
position ${\bm i} =(i_x, i_y)$ and the three directions $d=1,2,3$, corresponding to horizontal, vertical and diagonal wavelets respectively.
The corresponding index set $\Lambda$ is
$\Lambda = \left\{ \lambda=(j,i_x,i_y,d),j=0,...,J-1; i_x,i_y=0...2^j
-1,\right.$ $\left. d=1,2,3 \right\}\text{.}$ 
Due to orthogonality the wavelet coefficients are given by $\widetilde \omega_\lambda  =
\left< \omega, \psi_\lambda \right>$, where 
%{\bf
$\langle \cdot, \cdot \rangle$ denotes the $L^2$-inner product defined as 
$\langle f , g \rangle = \int f(x,y) g(x,y) dx dy$.
%}
The wavelet coefficients measure fluctuations of $\omega$ at scale $2^{-j}$ 
around the position $\bm i$, in one of the three directions $d$. Here
a Coifman 30 wavelet is used, which is orthogonal and has 10 vanishing
moments \cite{Mallat98} ($\int x^n \psi(x) dx = 0$ for $n =0,...9$).

The CVE algorithm can be summarized in the following three
step procedure:

\begin{itemize}
\item
{\em Decomposition:}
compute the wavelet coefficients $\widetilde \omega_\lambda$ using the fast 
wavelet transform\cite{Farge92}.
\item
{\em Thresholding:}
apply the thresholding function $\rho_\varepsilon$ to the wavelet
coefficients $\widetilde \omega_\lambda$,
thus discarding the coefficients with absolute values smaller than the threshold $\varepsilon$.
\item
{\em Reconstruction:}
reconstruct the coherent vorticity field $\omega_C$ from the thresholded wavelet coefficients using the fast inverse wavelet transform.
\end{itemize}
The incoherent vorticity field is obtained by simple subtraction, i.e., $\omega_I = \omega - \omega_C$.

The thresholding function is given by
\begin{equation}
\rho_\varepsilon(a)=\left\{\substack{a \text{ if } |a|>\varepsilon \\ 0 
\text{ if } |a|\leq \varepsilon}\right. \label{for:rho},
\end{equation}
where $\varepsilon$ denotes the threshold, %which depends on the enstrophy $Z$ and on the resolution $N$ only.
%Here we use the universal threshold 
\begin{equation}
\varepsilon= \sqrt{4 Z \ln N}\label{for:donoho},
\end{equation}
where $Z=\frac{1}{2}\langle\omega,\omega\rangle$ is the enstrophy
(which corresponds to half of the variance of the vorticity fluctuations) and $N$ the resolution. This threshold value allows for optimal denoising in a minmax sense, assuming the noise to be additive, Gaussian and white\cite{Pellegrino2001}. 

In summary, this decomposition yields $\omega =  \omega_C +  \omega_I$.
Due to orthogonality we have $\langle \omega_C ,  \omega_I \rangle = 0$
and hence it follows that enstrophy is conserved, i.e., $Z = Z_C + Z_I$. 
Let us mention that the computational cost of the Fast Wavelet Transform (FWT) is
of $O (N)$ \cite{Farge92}. 
%where N denotes the total number of grid points \cite{Farge92}.

\subsection{Compression rates}

\begin{table}
\caption{\label{Table1}Compression rate ($\%$ of coefficients retained), retained energy $E=\frac{1}{2}\left<\phi~\nabla^2\phi\right>$, enstrophy $Z=\frac{1}{2}\left<\omega^2\right>$, and radial flux $\Gamma_r$, after applying the CVE filter to the vorticity field of the quasi-hydrodynamic and quasi-adiabatic 2D drift-wave turbulence simulations.\\
}
\begin{tabular}{c c c c c}
\hline
\hline
          &Compr. (\%) & $E$ (\%)  & $Z$ (\%) & $\Gamma_r$ (\%) \\
\hline
Quasi-hydrodynamic ($c$=0.01) &1.3&99.9&97&99\\
Quasi-adiabatic ($c$=0.7)    &1.8&99.0 &93&98\\
\hline
\hline
\end{tabular}
\end{table}

The results of the extraction are displayed in table \ref{Table1}. The compression rate is in both cases very significant: for the quasi-hydrodynamic case, $1.3\%$ of the modes retain more than $99.9\%$ of the energy and $97\%$ of the enstrophy. For the quasi-adiabatic case, $1.8\%$ of the modes  retain $99.0\%$ of the energy and $93\%$ of the enstrophy. The contribution of the coherent vorticity to the radial flux is also given in table \ref{Table1}. The coherent modes, which contain most of the energy and enstrophy, are responsible for $99\%$ of the radial particle density flux $\Gamma_r$ in the quasi-hydrodynamic case, and for $98\%$ of $\Gamma_r$ in the quasi-adiabatic case. In other words, $\Gamma_r$ is almost exclusively carried by the coherent structures.

\subsection{Wavenumber spectra and probability density functions}

\begin{figure*}
\setlength{\unitlength}{1.\textwidth}
%\vspace{0.5cm}
%\includegraphics[width=0.25\unitlength]{fig2a.eps}\includegraphics[width=0.25\unitlength]{fig2b.eps}
%FIGURES/PDFvort512hydro.eps}
%{FIGURES/PDFvort512quasi.eps}
%\includegraphics[width=0.25\unitlength]{fig2c.eps}\includegraphics[width=0.25\unitlength]{fig2d.eps}
%{FIGURES/SPCvort512hydro.eps}
%{FIGURES/SPCvort512quasi.eps}
\includegraphics[width=1\unitlength]{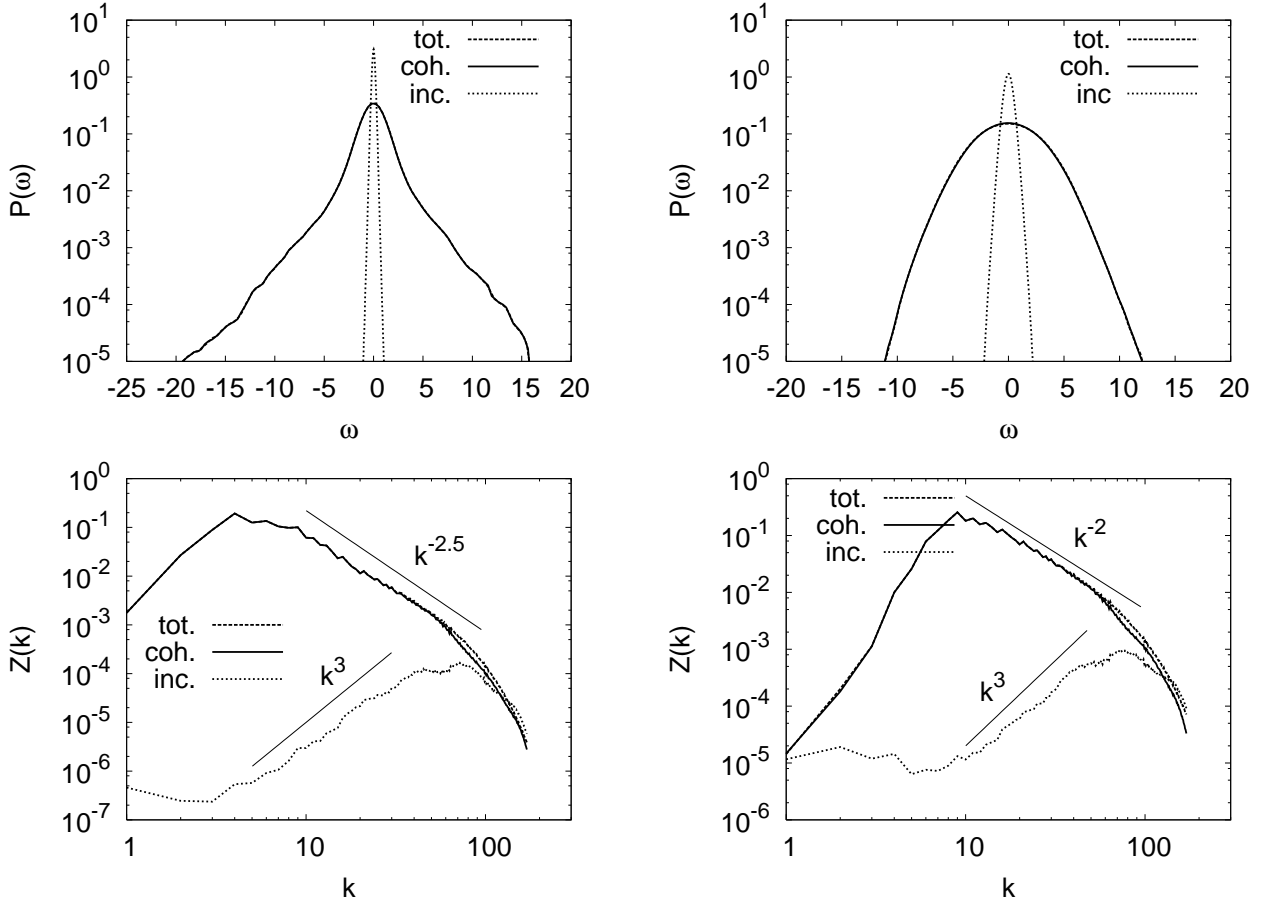}
\caption{Top: PDF of the vorticity. Bottom: Fourier spectrum of the
  enstrophy versus wavenumber. Left: quasi-hydrodynamic case. Right: quasi-adiabatic case. Dashed line: total field, solid line: coherent part, dotted
  line: incoherent part. Note that the coherent contribution (solid)
  superposes the total field (dashed), which is thus hidden under the
  solid line in all four figures. 
%{\bf 
The straight lines indicating
  power laws are plotted for reference.
%}
\label{PDFvort512}}
\end{figure*}

\begin{figure*}
\setlength{\unitlength}{1.0\textwidth}
%\vspace{0.5cm}
%\includegraphics[width=0.5\unitlength]{fig3a.eps}\includegraphics[width=0.5\unitlength]{fig3b.eps}
%{/scratch/BOS/SHINPEI/SCATTER/ZOOMHYDRO/scatter_tot_hydro.eps}\hspace{-0.cm}
%{/scratch/BOS/SHINPEI/SCATTER/ZOOMQUASI/scatter_tot_quasi.eps}\hspace{-0.cm}\\
%\includegraphics[width=0.5\unitlength]{fig3c.eps}\includegraphics[width=0.5\unitlength]{fig3d.eps}
%{/scratch/BOS/SHINPEI/SCATTER/ZOOMHYDRO/scatter_inc_hydro.eps}\hspace{-0.cm}
%{/scratch/BOS/SHINPEI/SCATTER/ZOOMQUASI/scatter_inc_quasi.eps}\hspace{-0.cm}
\includegraphics[width=1\unitlength]{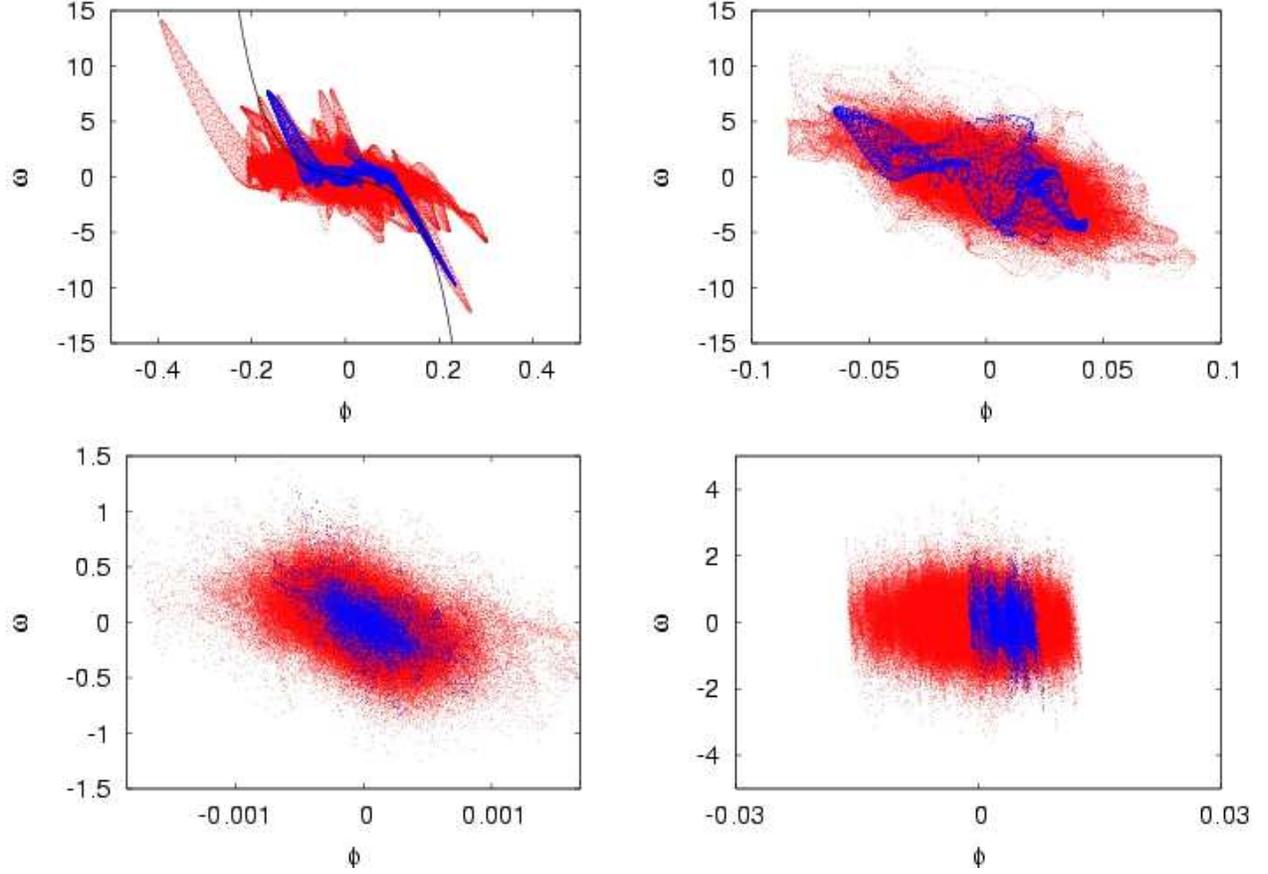}
\caption{(Color online) Scatter-plot of vorticity against electrostatic potential for the coherent part (top) and incoherent part (bottom). Left quasi-hydrodynamic case, right quasi-adiabatic case.  The light grey (red online) dots correspond to the total field, the dark grey (blue online) dots to the dipoles we have selected in Fig. 1.    \label{scatter}}
\end{figure*}

Spectra and probability density functions (PDF), averaged over $512$
realizations during the time interval $100 < t \leq 612$, are shown in
Figure \ref{PDFvort512} for the total, coherent and incoherent
vorticity. The PDF of the total and coherent quasi-hydrodynamic
vorticity is far from Gaussian and slightly skewed, while the
quasi-adiabatic vorticity  is much closer to Gaussianity. In both
cases, the variance of the incoherent part is much smaller than the
variance of the coherent part, which has the same PDF as the
total.
%{\bf 
For the quasi-hydrodynamic case, the coherent part retains $97\%$ of
the variance of the vorticity fluctuations and therefore also
$97\%$ of the total enstrophy $Z$, with $Z=1.4$.  For the quasi-adiabatic case, the coherent
part retains $93\%$ of
the variance of the vorticity fluctuations and hence
$93\%~Z$, with $Z=3.4$.
%enstrophy is $93\%~Z$ with $Z=3.4$.
%incoherent part $Z_I=0.04$. For the quasi-adiabatic case $Z=3.4$ and the incoherent part $Z_I=0.25$.
%}  
A similar result is observed in the enstrophy spectrum 
%{\bf $Z(k)$, 
computed from the Fourier transform of the vorticity field, averaged over
wavenumber shells of radius $|\bm k |$, the wavenumber.
%} 
The total and
coherent enstrophy are the same all over the inertial range and 
at the highest wavenumbers, in the dissipation range, the incoherent
part contributes to the spectral enstrophy density. Both coherent and
incoherent contributions are spread all over the spectral range, but
they present different
spectral slopes in the inertial range and therefore different spatial correlations. From the integral wavenumber to the dissipation wavenumber, a
 negative slope for the coherent contribution, corresponding to long range spatial correlations, is observed. The incoherent part shows a positive slope with a power-law dependence close to $k^3$ in the inertial range. This corresponds to an equipartition of kinetic energy in two dimensions. A similar result was obtained in three-dimensional isotropic Navier-Stokes turbulence\cite{Pellegrino2001}.
%}

%These findings are in agreement with the results found in 2d and 3d hydrodynamic turbulence 
%\cite{FSK99}. 
%}

\subsection{Scatter-plots}

We show in figure \ref{scatter} scatter-plots of the vorticity versus
the electrostatic potential corresponding to the fields in figure
\ref{VortFields}. Both the total part and the incoherent part are
shown. Since the coherent part is almost identical to the total part,
it has been omitted. Also shown, superposed on the same figures, is
the scatter-plot corresponding to the zoom on the dipolar structures
indicated by a white frame in figure \ref{VortFields}. In the freely
decaying hydrodynamic case, $c=0$, Joyce and Montgomery
\cite{Joyce1973} showed that
a functional relation $\phi(\omega)=\alpha \sinh(\beta\omega)$ should
be expected, corresponding to a final state of decay depleted from
nonlinearity. 
%{\bf
The parameters $\alpha$ and $\beta$ are Lagrangian multipliers,
necessary for maximizing the entropy under constraints. The value
$1/\beta$ can be associated with a (negative) temperature \cite{Joyce1973}. 
%}
Depletion from nonlinearity corresponds to steady solutions of the Euler equation, $\left[\omega,\phi\right]=0$, implied by the existence of a functional relation $\phi(\omega)$. Indeed drift-wave turbulence contains an internal instability which prevents the flow from  decaying. This forcing is present in both cases considered here and a \emph{sinh-Poisson} relation cannot be expected \emph{a priori} for the global flows. Moreover, the two-field model [equations (\ref{hw1}) and (\ref{hw2})] contains two nonlinearities, first the polarization-drift nonlinearity in the vorticity equation, second the $E\times B$ nonlinearity in the density equation. The latter disappears in the adiabatic limit as $n$ and $\phi$ are in phase, which corresponds to a linear functional relationship. In figure \ref{scatter}, a local depletion of the polarization-drift nonlinearity is seen for the quasi-hydrodynamic case. The scatter-plot of $\phi - \omega$, corresponding to the dipolar structure, that is indicated by a white frame in figure \ref{VortFields} (left), is close to a \emph{sinh-Poisson} relation (solid black curve) in spite of the presence of the forcing term.
%Nevertheless, in figure \ref{scatter}, the scatter plot corresponding to the dipolar structure that is indicated by a white frame in figure \ref{VortFields}, is close to this relation (solid black curve) for the quasi-hydrodynamic case, which corresponds to a local depletion of nonlinearity. 
In the quasi-adiabatic case the dipolar structure, that is indicated by a white frame in figure \ref{VortFields} (right), does not exhibit such a functional relation. In the incoherent parts (Figure \ref{scatter}, bottom) no functional relation can be distinguished, which confirms that the incoherent part does not contain any structure, for both quasi-hydrodynamic and quasi-adiabatic cases.

\subsection{Strain versus vorticity}

A question is now how to quantitatively distinguish between the structures in both cases. Intuitively it can be inferred that  different regions of high vorticity in the quasi-adiabatic case involve strong mutual shearing which strongly limits their lifetime and the chance to reach a functional relation $\phi(\omega)$.  Koniges \etal \cite{Koniges1992} determined the lifetime of individual eddies compared to the eddy-turnover time $\tau_{over}$, i.e. the time it takes for a fluid element in an eddy to make a $2\pi$ rotation. They estimated the lifetime of the quasi-hydrodynamic eddies to be approximately 10 $\tau_{over}$, and the lifetime of the adiabatic eddies (for $c=2.0$) approximately $\tau_{over}$.  As mentioned in their paper, this measure is quite subjective and very time-consuming, especially if a full PDF of the lifetimes is to be obtained. Here we propose a simpler way to distinguish the coherent structures for the different regimes.

In fluid turbulence the Weiss criterion $Q$ \cite{Weiss1991} is a local measure of the  strain compared to the vorticity for a 2D velocity field. The Weiss field is defined as:
\begin{equation}
Q=\frac{1}{4}\left(\sigma^2-\omega^2\right),
\end{equation}
with 
\begin{equation}
\sigma^2=\left(\frac{\partial u}{\partial x}-\frac{\partial v}{\partial
  y}\right)^2+\left(\frac{\partial u}{\partial y}+\frac{\partial v}{\partial x}\right)^2.
\end{equation}
$u$ and $v$ are two orthogonal components of the velocity vector.
  The Weiss criterion was proposed to identify coherent structures, but it may lead to ambiguous results because the underlying assumption that the velocity gradient varies slowly with respect to the vorticity gradient is not generally valid \cite{Basdevant1994}. We here apply the same criterion to drift-wave turbulence \cite{Pedersen1996,Annibaldi2002,Naulin1997} but not to identify coherent structures (this being done by the CVE method), but to distinguish between the quasi-hydrodynamic and quasi-adiabatic cases. 

\begin{figure}
\setlength{\unitlength}{1.\textwidth}
%\vspace{0.5cm}
\includegraphics[width=0.5\unitlength]{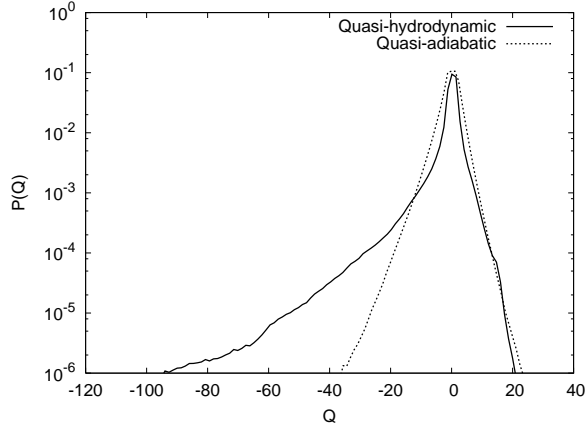}
%{/scratch/BOS/SHINPEI/VISU/PDFQ512.eps}~~
\caption{PDF of the Weiss field $Q$ for the quasi-hydrodynamic and quasi-adiabatic velocity fields. \label{PDFQ}}
\end{figure}

The PDF of the Weiss field (Fig.\ref{PDFQ}) reveals that it is its skewness that differentiates best the two fields. Indeed, it
is more skewed towards negative $Q$ for the quasi-hydrodynamic case
than for the quasi-adiabatic case: the skewness is $-11$ for the
former, compared to $-2$ for the latter. The PDF shows thus that in the 
quasi-hydrodynamic case the probability to find rotationally dominated 
regions is larger, and the rotation exhibits much larger values, than 
in the quasi-adiabatic case. The variance of $Q$ is comparable
for the two cases ($5$ and $4$ for the quasi-hydrodynamic case and the
quasi-adiabatic case, respectively). 
%{\bf
The skewness of the Weiss field $Q$ appears to be a good quantitative measure
to distinguish between the two cases studied in the present
work. In further studies it can be investigated, whether this
measure can be used to identify coherence in different types of turbulent flows.%}
%We propose the skewness of the
%Weiss field as a quantitative measure to distinguish between the two
%cases. 
%{\bf
%The large negative values of $Q$ are related to the existence of the
%coherent vortical structures ... to be completed 
%}

\section{Conclusion and perspectives}

In conclusion, we have applied the \emph{Coherent Vortex Extraction} method to dissipative drift-wave turbulence. The results show that we can identify the essential degrees  of freedom (less than $2\%$) responsible for the nonlinear dynamics and transport. The coherent modes contain almost all the energy and enstrophy and contribute to more than $98\%$ of the radial flux. 

%{\bf 
Evaluating the scatter-plot of the vorticity versus the electrostatic
potential, %for the $E\times B$-motion, 
it is shown that the coherent
structures in the quasi-hydrodynamic case are close to a state of
local depletion of polarization-drift nonlinearity. In contrast, this is not the case for the quasi-adiabatic regime, where nonlinearity remains active and no \emph{sinh}-functional relation between vorticity and electrostatic potential is observed.
%the mutual shearing of the coherent structures
%inhibits such a depetion of non-linearity and the
%is
%therefore not observed. 
%}
This depletion of nonlinearity in the quasi-hydrodynamic regime may 
explain the failure of the
quasi-linear estimate of the radial flux\cite{Koniges1992}. The skewness of the Weiss field yields a
quantitative measure for the difference in nonlinear behavior of the
coherent structures between the quasi-hydrodynamic and quasi-adiabatic
cases.

The wavelet transforms, or the proper orthogonal decomposition (POD), may become very useful to denoise particle-in-cell simulations of plasma turbulence\cite{Gassama2006}. A comparison of the performance of the POD and CVE method is currently undertaken and will be reported in a future paper.

\section*{Acknowledgments}

\vfill
Lionel Larchev\^eque is acknowledged for supplying and helping with a
routine to compute the Weiss field.  
%{\bf 
Wendel Horton is acknowledged for comments on the manuscript.
%}
This work was supported by the \emph{Agence Nationale de la Recherche} under the contract \emph{M\'ethodes Multi-\'echelles pour l'analyse et la simulation num\'erique en Turbulence Fluide et Plasma} (M2TFP).

%\bibliography{../../biblio}

\end{document}